\documentclass[12pt]{article}

\usepackage{amssymb}
\usepackage{amsfonts}

\begin{document}

\title{Quantum transport in degenerate systems}

\author{S.V. Kozyrev}

\maketitle

\centerline{Steklov Mathematical Institute of Russian Academy of Sciences, Moscow, Russia}

\begin{abstract}
Transport in nonequilibrium degenerate quantum systems is investigated. Transfer rate depends on parameters of the system. In this paper we investigate dependence of the flow (transfer rate) on the angle between ''bright'' vectors (which define interaction of the system with the environment). We show that in some approximation for the system under investigation the flow is proportional to cosine squared of the angle between ''bright'' vectors. Earlier in paper by the author it was shown that in this degenerate quantum system excitation of non--decaying quantum ''dark'' states is possible, moreover the effectiveness of this process is proportional to sine squared of the angle between ''bright'' vectors (this phenomenon was discussed as a possible model of excitation of quantum coherences in quantum photosynthesis). Thus quantum transport and excitation of dark states are competing processes, ''dark'' states can be considered as a result of leakage of quantum states in quantum thermodynamic machine which performs the quantum transport.
\end{abstract}

\section{Introduction}

In the present paper we consider a three--level quantum system with one degenerate energy level. This system interacts with nonequilibrium environment given by three Bose quantum fields (or {\it reservoirs}) in Gibbs states with different temperatures. This system gives an example of quantum thermodynamic machine (interaction of the system with nonequilibrium environment generates thermodynamic cycle related to quantum transport).  Moreover interactions of the system with different reservoirs have different forms (are related to non--parallel ''bright'' vectors, see below). We find nonequilibrium stationary state for this system and compute the rate of transfer (of the {\it flow}) between the energy levels. We find the dependence of the transfer rate on properties of the model --- states of the reservoirs (intensities of the fields) and angle between the ''bright'' vectors (related to interactions with different reservoirs). In the first approximation the flow is proportional to $\cos^2\alpha$ where $\alpha$ is the angle between the ''bright'' vectors.

Let us note that the previous paper of the author \cite{darkstates} investigated possibility of manipulation of quantum states in analogous quantum system, where one of the reservoirs of the nonequilibrium environment also contains a coherent component (laser field). It was shown that in this setup one can generate non--decaying quantum ''dark'' states (dark states were studied in quantum optics \cite{Scully}, \cite{dark-state} in relation to induced transparency, quantum memory and light stopping). These states can be observed by repeated interaction with the coherent field (as in the experiments on photonic echo, see below). The corresponding response for this interaction with coherent field was proportional to $\sin^2\alpha$ where $\alpha$ is the mentioned above angle between bright vectors.

In the present paper for the system under consideration we study the transfer rate in the stationary state in the case of non--zero angle between the bright states and investigate dependence of the transfer rate on this angle (in the first approximation the flow is proportional to $\cos^2\alpha$) and other parameters of the system. This dependence is the main result of the present paper.

Thus in the present model quantum transport and generation of non--decaying quantum ''dark'' states are competing processes (schematically this is described by the formula $\cos^2\alpha+\sin^2\alpha=1$).

From the physical point of view the model under consideration describes quantum photosynthesis, i.e. the combination of processes of absorption of photons by chromophores with creation of excitons (excited electronic states of chromophores), transport of excitons to the reaction center and absorption of excitons in the reaction center (these three processes form thermodynamic cycle of a quantum thermodynamic machine). Three reservoirs describe interaction of the system with light (photons), with phonons (vibrations of the protein matrix), interaction with the third reservoir describes the sink of excitons  (absorption of excitons in the reaction center). Temperatures of three reservoirs are different  --- it is natural to take the temperature of light equal to 6000K, and temperatures of phonons and sink equal to 300K (room temperature).

In paper \cite{darkstates} it was proposed to use the above mentioned quantum dark states for description of the effect of quantum photosynthesis which attracted a lot of attention in recent years \cite{Engel}, \cite{SFOG}, \cite{Mohseni}, \cite{Qbio}. This effect is related to experimental observation of photonic echo in photosynthetic systems, which proves the existence of quantum coherences with long lifetime for these systems. In the present paper we study the influence of these states to the rate of quantum transport of excitons in photosynthetic systems. The obtained results show that the effect of quantum photosynthesis (existence of the photonic echo) can be considered as the result of leakage of quantum states in poorly developed quantum thermodynamic machine.

For investigation of described model we use method of quantum stochastic limit \cite{AcLuVo} which allows to describe phenomena of relaxation, decoherence and transfer in open quantum systems, see also \cite{notes}. In \cite{OhyaVolovich} applications of quantum methods in computations and biology were considered. Photosynthetic complexes give example of complex physical systems, see \cite{complex_review} for review of application of ultrametric methods to complex systems in physics and biology.

Effect of quantum photosynthesis is related to excitation of quantum states with long lifetime. Different approaches to control of coherence in quantum systems and manipulation of quantum states were discussed in \cite{TruVol}, \cite{lambdaatom}, \cite{Vol}, \cite{Vol-2}, \cite{Pechen}, \cite{PechenTru}, \cite{Pechen1}, \cite{Pechen2}, \cite{Amosov}.  In \cite{Holevo1}, \cite{Shirokov} problems of quantum information theory were considered, in particular capacity of quantum channels.

In \cite{Arefeva} a new approach to quantum photosynthesis based on the so called holographic approach was considered. Holographic approach earlier was applied in high energy physics \cite{Arefeva1}, \cite{Arefeva2}, \cite{Arefeva3}.

The structure of the present paper is as follows. In Section 2 we describe system interacting with nonequilibrium environment in the form of three reservoirs (quantum Bose fields), interaction of the system with the environment depends on ''bright'' photonic and phononic vectors. In Section 3 generators of quantum dissipative dynamics of reduced density matrix of the system are built with application of the quantum stochastic limit approach.  In Section 4 we construct the stationary state of the system for the case of parallel photonic and phononic bright vectors  and compute the flow (rate of quantum transfer) in this stationary state, and discuss dependence  of the transfer rate on the states of reservoirs. In Section 5 these results are generalized for the case of non--parallel bright vectors and dependence of the flow on the angle between photonic and phononic bright vectors is discussed.

\section{Description of the model}

We consider a quantum system with three energy levels $\varepsilon_0<\varepsilon_1<\varepsilon_2$, two of levels are non--degenerate, and Hamiltonian
\begin{equation}\label{H_S}
H_S=\varepsilon_0 |0\rangle\langle 0|+ \varepsilon_1 |1\rangle\langle 1|+\varepsilon_2\sum_{j=2}^{N} |j\rangle\langle j|.
\end{equation}

This Hamiltonian describes photosynthetic system in one exciton approximation in the so called global basis, $|0\rangle$ is the state without excitons, $|1\rangle$ is the state ''exciton in the reaction center'' (or {\it sink}), $|j\rangle$ correspond to one exciton states of chromophores.

The system interacts with three reservoirs in temperature states. Transitions between energy levels $\varepsilon_0$ and $\varepsilon_2$ (in particular creation of excitons) are related to interaction with light (temperature reservoir with $\beta_{\rm em}^{-1}=6000K$), transitions between levels $\varepsilon_2$ and $\varepsilon_1$ (transfer of excitons to the reaction center) are related to interaction with vibrations of the protein matrix  (temperature reservoir with $\beta_{\rm ph}^{-1}=300K$), and transitions between levels $\varepsilon_1$ and $\varepsilon_0$ (absorption of excitons in the reaction center) are related to interaction with the sink reservoir with temperature $\beta_{\rm sink}^{-1}=300K$.

Dynamics of the model runs in the Hilbert space (tensor product of Hilbert spaces of the system and reservoirs)
$$
{\cal H}={\cal H}_S\otimes {\cal H}_{\rm em}\otimes {\cal H}_{\rm ph}\otimes {\cal H}_{\rm sink}.
$$

Thus we have three reservoirs described by Hamiltonians of quantum Bose fields $H_{\rm em}$ (electromagnetic field), $H_{\rm ph}$ (phonons, or vibrations of the protein matrix), $H_{\rm sink}$ (this reservoir describes absorption of excitions in the reaction center), Hamiltonians of reservoirs have the form
$$
H_R=\int_{\mathbb{R}^3} \omega_R(k)a^{*}_R(k) a_R(k) dk,
$$
where index $R={\rm em},\, {\rm ph},\, {\rm sink}$ enumerate reservoirs, $\omega_R$ is dispersion of the Bose field $a_R$, commutation relations for the field have the form
$$
[a_R(k),a^{*}_R(k')]=\delta(k-k').
$$

Each of reservoirs is in the temperature state, i.e. gaussian non--squeezed mean zero state with non--zero quadratic correlation functions of the form
$$
\langle a^{*}_{R}(k)a_R(k') \rangle=N_R(k)\delta(k-k').
$$
Here $N_R(k)$ (number of field quanta with wave number $k$ for the state of the field) is equal to (in the temperature state with inverse temperature $\beta_R$)
\begin{equation}\label{temperature}
N_R(k)={1\over{e^{\beta_R\omega_R(k)}-1}}.
\end{equation}

Total Hamiltonian of the model is equal to
\begin{equation}\label{H_total}
H=H_S+H_{\rm em}+H_{\rm ph}+H_{\rm sink}+ \lambda\left( H_{I,{\rm em}}+H_{I,{\rm ph}}+H_{I,{\rm sink}}\right),
\end{equation}
where Hamiltonians of interaction of the system with reservoirs $H_{I,{\rm em}}$, $H_{I,{\rm ph}}$, $H_{I,{\rm sink}}$ have the form (\ref{H_em}), (\ref{H_ph}), (\ref{H_sink}) correspondingly. In these formulae $g_{R}(k)$, $R={\rm em},\, {\rm ph},\, {\rm sink}$ are form factors of interactions (complex valued functions).

Interaction of the system with light is described by interaction Hamiltonian
\begin{equation}\label{H_em}
H_{I,{\rm em}} = A_{\rm em}|\chi\rangle\langle 0| + A^{*}_{\rm em}|0\rangle\langle \chi|,\qquad A^{*}_{\rm em}=\int_{\mathbb{R}^3} g_{\rm em}(k)a^{*}_{\rm em}(k)  dk,
\end{equation}
where bright photonic vector $\chi$ belongs to energy level $\varepsilon_2$ (upper degenerate energy level of the system), $g_{\rm em}(k)$ is form factor of interaction.

Exciton transport is related to interaction with phonons
\begin{equation}\label{H_ph}
H_{I,{\rm ph}} = A_{\rm ph}|\psi\rangle\langle 1| + A^{*}_{\rm ph}|1\rangle\langle \psi|,\qquad A^{*}_{\rm ph}=\int_{\mathbb{R}^3} g_{\rm ph}(k)a^{*}_{\rm ph}(k)  dk,
\end{equation}
here bright phononic vector $\psi$ has energy $\varepsilon_2$ (belongs to the same energy level as vector $\chi$).

Vectors $\psi$ and $\chi$ belong to the same space of excitons on chromophores (degenerate level with energy $\varepsilon_2$). Principal feature of the model under investigation is that vectors $\psi$ and $\chi$ are non--parallel. Since interactions of the system with photons and phonons are physically different the corresponding bright vectors are not necessarily parallel. We denote $\alpha$ the angle between vectors $\chi$ and $\psi$, i.e.
$$
|\langle\psi,\chi \rangle|=\cos\alpha \|\psi\|\|\chi\|.
$$

Absorption of excitons in the reaction center is described by interaction with the sink reservoir 
\begin{equation}\label{H_sink}
H_{I,{\rm sink}} = A_{\rm sink}|1\rangle\langle 0| + A^{*}_{\rm sink}|0\rangle\langle 1|,\qquad A^{*}_{\rm sink}=\int_{\mathbb{R}^3} g_{\rm sink}(k)a^{*}_{\rm sink}(k)  dk.
\end{equation}

\section{Generators of dynamics}

For investigation of dynamics of the system we use the approach of the quantum stochastic limit \cite{AcLuVo}. Generalization of generators of dynamics of the stochastic limit for the case of quantum many particle systems (in particular degenerate) can be found in \cite{notes}, \cite{tmf2014}. In this limit dynamics of the reduced density matrix of the system interacting with environment is described by some particular generators in the Lindblad form, see below.

For the model considered here dynamics of the reduced density matrix of the system interacting with three reservoirs is generated by the sum of three Lindblad generators
\begin{equation}\label{total_gen}
\frac{d}{dt}\rho(t)=\left(\theta_{\rm em}+\theta_{\rm ph}+\theta_{\rm sink}\right)(\rho(t)).
\end{equation}

The photonic generator describes generation of excitons 
\begin{equation}\label{theta_em}
\theta_{\rm em}(\rho)=
\|\chi\|^2\biggl[2\gamma^{-}_{{\rm re,em}}
\left(
\langle \widetilde{\chi}|\rho|\widetilde{\chi}\rangle |0\rangle\langle 0|
-{1\over 2}
\{\rho,|\widetilde{\chi}\rangle\langle \widetilde{\chi}|\}\right)
-i\gamma^{-}_{{\rm im,em}} [\rho,|\widetilde{\chi}\rangle\langle \widetilde{\chi}|]+
$$
$$
+
2\gamma^{+}_{{\rm re,em}}
\left(\langle 0| \rho |0\rangle |\widetilde{\chi}\rangle\langle \widetilde{\chi}|
-{1\over 2}
\{\rho,|0\rangle\langle 0| \}\right)
+i\gamma^{+}_{{\rm im,em}} [\rho,|0\rangle\langle 0|]\biggr].
\end{equation}

Here normed bright photonic vector has the form
$$
|\widetilde{\chi}\rangle=\frac{|\chi\rangle}{\|\chi\|}.
$$

The constants $\gamma$ are given by (\ref{Regg+})--(\ref{Imgg-}), Bohr frequency $\omega_R$ for $R={\rm em}$ equals to $\omega_{\rm em}=\varepsilon_2-\varepsilon_0$, function $N_{\rm em}(k)$ is given by (\ref{temperature}), $\beta_{\rm em}^{-1}=6000K$ for the sun light.
\begin{equation}\label{Regg+}
\gamma^{+}_{{\rm re},R}=\pi\int |g_R(k)|^2\delta(\omega_R(k)-\omega_R)N_R(k)dk,
\end{equation}
\begin{equation}\label{Regg-}
\gamma^{-}_{{\rm re},R} =\pi\int |g_R(k)|^2\delta(\omega_R(k)-\omega_R)(N_R(k)+1)dk,
\end{equation}
\begin{equation}\label{Imgg+}
\gamma^{+}_{{\rm im},R}=-\,{\rm P}\,\int |g_R(k)|^2{1\over\omega_R(k)-\omega_R}N_R(k)dk,
\end{equation}
\begin{equation}\label{Imgg-}
\gamma^{-}_{{\rm im},R} =-\,{\rm P}\,\int |g_R(k)|^2{1\over\omega_R(k)-\omega_R}(N_R(k)+1)dk.
\end{equation}
Here ${\rm P}$ is the Cauchy principal value.

Transport of excitons is described by the phononic generator, here $R={\rm ph}$, Bohr frequency has the form $\omega_{\rm ph}=\varepsilon_2-\varepsilon_1$ with temperature $\beta_{\rm ph}^{-1}=300K$
\begin{equation}\label{theta_ph}
\theta_{\rm ph}(\rho)=
\|\psi\|^2\biggl[2\gamma^{-}_{{\rm re,ph}}
\left(
\langle \widetilde{\psi}|\rho|\widetilde{\psi}\rangle |1\rangle\langle 1|
-{1\over 2}
\{\rho,|\widetilde{\psi}\rangle\langle \widetilde{\psi}|\}\right)
-i\gamma^{-}_{{\rm im,ph}} [\rho,|\widetilde{\psi}\rangle\langle \widetilde{\psi}|]+
$$
$$
+
2\gamma^{+}_{{\rm re,ph}}
\left(\langle 1| \rho |1\rangle |\widetilde{\psi}\rangle\langle \widetilde{\psi}|
-{1\over 2}
\{\rho,|1\rangle\langle 1| \}\right)
+i\gamma^{+}_{{\rm im,ph}} [\rho,|1\rangle\langle 1|]\biggr].
\end{equation}
The normed bright phononic vector is
$$
|\widetilde{\psi}\rangle=\frac{|\psi\rangle}{\|\psi\|}.
$$

Absorption of excitons is given by the sink generator, where $R={\rm sink}$, Bohr frequency is $\omega_{\rm sink}=\varepsilon_1-\varepsilon_0$ with temperature $\beta_{\rm sink}^{-1}=300K$
\begin{equation}\label{theta_sink}
\theta_{\rm sink}(\rho)=
2\gamma^{-}_{{\rm re,sink}}
\left(
\langle 1|\rho|1\rangle |0\rangle\langle 0|
-{1\over 2}
\{\rho,|1\rangle\langle 1|\}\right)
-i\gamma^{-}_{{\rm im,sink}} [\rho,|1\rangle\langle 1|]+
$$
$$
+2\gamma^{+}_{{\rm re,sink}}
\left(
\langle 0|\rho|0\rangle |1\rangle\langle 1|
-{1\over 2}
\{\rho,|0\rangle\langle 0|\}\right)
+i\gamma^{+}_{{\rm im,sink}} [\rho,|0\rangle\langle 0|].
\end{equation}

Constants $\gamma$ in (\ref{theta_ph}), (\ref{theta_sink}) are described by formulae (\ref{Regg+}), (\ref{Regg-}), (\ref{Imgg+}), (\ref{Imgg-}) with Bohr frequencies, dispersions of the fields, form factors of interactions and temperatures of the reservoirs corresponding to phonons and sink.

\section{Transfer rate for $\alpha=0$}

In present Section we compute the stationary state of the system interacting with nonequilibrium environment and the rate of quantum transport for the case $\alpha=0$ (when bright photonic and phononic vectors $\chi$ and $\psi$ are parallel). Flows in nonequilibrium quantum systems (including the case of several reservoirs) were studied in particular in \cite{Flows}, \cite{Imafuku}, \cite{Chi}.

In this case we can consider dynamics of the reduced density matrix of the system given by equation (\ref{total_gen}) in the subspace of diagonal matrices of the form
\begin{equation}\label{bright}
\rho=\rho_{00}|0\rangle\langle 0|+\rho_{11}|1\rangle\langle 1|+\rho_{\psi\psi}|\widetilde{\psi}\rangle\langle \widetilde{\psi}|,
\end{equation}
$$
\rho_{00}+\rho_{11}+\rho_{\psi\psi}=1.
$$

In this subspace we have quantum transfer of excitons, in particular there exists nonequilibrium stationary state with non--zero flow. This stationary state has the form 
\begin{equation}\label{rho22}
\rho_{\psi\psi}=\frac{
\gamma_{{\rm re},{\rm em} }^{+}\gamma_{{\rm re},{\rm ph} }^{+}\|\chi\|^2\|\psi\|^2+
\gamma_{{\rm re},{\rm em} }^{+}\gamma_{{\rm re},{\rm sink} }^{-}\|\chi\|^2+
\gamma_{{\rm re},{\rm ph} }^{+}\gamma_{{\rm re},{\rm sink} }^{+}\|\psi\|^2}{\Delta};
\end{equation}
\begin{equation}\label{rho11}
\rho_{11}=\frac{
\gamma_{{\rm re},{\rm em} }^{+}\gamma_{{\rm re},{\rm ph} }^{-}\|\chi\|^2\|\psi\|^2+
\gamma_{{\rm re},{\rm em} }^{-}\gamma_{{\rm re},{\rm sink} }^{+}\|\chi\|^2+
\gamma_{{\rm re},{\rm ph} }^{-}\gamma_{{\rm re},{\rm sink} }^{+}\|\psi\|^2
}{\Delta};
\end{equation}
\begin{equation}\label{rho00}
\rho_{00}=\frac{\gamma_{{\rm re},{\rm em} }^{-}\gamma_{{\rm re},{\rm ph} }^{+}\|\chi\|^2\|\psi\|^2+
\gamma_{{\rm re},{\rm em} }^{-}\gamma_{{\rm re},{\rm sink} }^{-}\|\chi\|^2+
\gamma_{{\rm re},{\rm ph} }^{-}\gamma_{{\rm re},{\rm sink} }^{-}\|\psi\|^2
}{\Delta}.
\end{equation}
where
\begin{equation}\label{Delta}
\Delta=
\left(\gamma_{{\rm re},{\rm ph} }^{+}\gamma_{{\rm re},{\rm em} }^{+}+
\gamma_{{\rm re},{\rm ph} }^{-}\gamma_{{\rm re},{\rm em} }^{+}+
\gamma_{{\rm re},{\rm ph} }^{+}\gamma_{{\rm re},{\rm em} }^{-}\right)\|\chi\|^2\|\psi\|^2+
$$ $$+
\left(\gamma_{{\rm re},{\rm ph} }^{+}\gamma_{{\rm re},{\rm sink} }^{+}+
\gamma_{{\rm re},{\rm ph} }^{-}\gamma_{{\rm re},{\rm sink} }^{+}+
\gamma_{{\rm re},{\rm ph} }^{-}\gamma_{{\rm re},{\rm sink} }^{-}\right)\|\psi\|^2+
$$ $$+
\left(\gamma_{{\rm re},{\rm em} }^{+}\gamma_{{\rm re},{\rm sink} }^{-}+
\gamma_{{\rm re},{\rm em} }^{-}\gamma_{{\rm re},{\rm sink} }^{+}+
\gamma_{{\rm re},{\rm em} }^{-}\gamma_{{\rm re},{\rm sink} }^{-}\right)\|\chi\|^2.
\end{equation}

Described stationary state for degenerate system under consideration is not unique, there exist also so called dark stationary states. For these states the rate of quantum transfer is equal to zero, see \cite{darkstates}, \cite{tmf2014}. There are also off--diagonal matrices on the system space, in the subspace of off--diagonal matrices decoherence (process of decay of quantum coherence) takes place.

Since reservoirs are in temperature states, formulae (\ref{Regg+}), (\ref{Regg-}) imply
$$
\frac{\gamma_{{\rm re},{\rm em} }^{+}}{\gamma_{{\rm re},{\rm em} }^{-}}=e^{-\beta_{\rm em}(\varepsilon_2-\varepsilon_0)},\qquad
\frac{\gamma_{{\rm re},{\rm ph} }^{+}}{\gamma_{{\rm re},{\rm ph} }^{-}}=e^{-\beta_{\rm ph}(\varepsilon_2-\varepsilon_1)},\qquad
\frac{\gamma_{{\rm re},{\rm sink} }^{+}}{\gamma_{{\rm re},{\rm sink} }^{-}}=e^{-\beta_{\rm sink}(\varepsilon_1-\varepsilon_0)}.
$$

Hence 
$$
\frac{\rho_{\psi\psi}}{\rho_{00}}=
\frac{\gamma_{{\rm re},{\rm em} }^{+}\gamma_{{\rm re},{\rm ph} }^{+}\|\chi\|^2\|\psi\|^2+
\gamma_{{\rm re},{\rm em} }^{+}\gamma_{{\rm re},{\rm sink} }^{-}\|\chi\|^2+
\gamma_{{\rm re},{\rm ph} }^{+}\gamma_{{\rm re},{\rm sink} }^{+}\|\psi\|^2}
{\gamma_{{\rm re},{\rm em} }^{-}\gamma_{{\rm re},{\rm ph} }^{+}\|\chi\|^2\|\psi\|^2+
\gamma_{{\rm re},{\rm em} }^{-}\gamma_{{\rm re},{\rm sink} }^{-}\|\chi\|^2+
\gamma_{{\rm re},{\rm ph} }^{-}\gamma_{{\rm re},{\rm sink} }^{-}\|\psi\|^2}=
$$
$$
=e^{-\beta_{\rm em}(\varepsilon_2-\varepsilon_0)}
\frac{\gamma_{{\rm re},{\rm em} }^{-}\gamma_{{\rm re},{\rm ph} }^{+}\|\chi\|^2\|\psi\|^2+
\gamma_{{\rm re},{\rm em} }^{-}\gamma_{{\rm re},{\rm sink} }^{-}\|\chi\|^2+
e^{\beta_{\rm em}(\varepsilon_2-\varepsilon_0)}e^{-\beta_{\rm ph}(\varepsilon_2-\varepsilon_1)}e^{-\beta_{\rm sink}(\varepsilon_1-\varepsilon_0)}
\gamma_{{\rm re},{\rm ph} }^{-}\gamma_{{\rm re},{\rm sink} }^{-}\|\psi\|^2}
{\gamma_{{\rm re},{\rm em} }^{-}\gamma_{{\rm re},{\rm ph} }^{+}\|\chi\|^2\|\psi\|^2+
\gamma_{{\rm re},{\rm em} }^{-}\gamma_{{\rm re},{\rm sink} }^{-}\|\chi\|^2+
\gamma_{{\rm re},{\rm ph} }^{-}\gamma_{{\rm re},{\rm sink} }^{-}\|\psi\|^2}.
$$

If the environment is equilibrium (i.e. inverse temperatures $\beta$ are equal for all reservoirs) we get
$$
e^{\beta_{\rm em}(\varepsilon_2-\varepsilon_0)}e^{-\beta_{\rm ph}(\varepsilon_2-\varepsilon_1)}e^{-\beta_{\rm sink}(\varepsilon_1-\varepsilon_0)}=1,
$$
thus $\rho_{\psi\psi}/\rho_{00}=e^{-\beta_{\rm em}(\varepsilon_2-\varepsilon_0)}$, i.e. the described above stationary state of the system is also equilibrium.

If temperatures of phonons and sink are equal $\beta_{\rm sink}=\beta_{\rm ph}$ (but not equal to temperature of phonons) we get
$$
\frac{\rho_{\psi\psi}}{\rho_{00}}=
e^{-\beta_{\rm em}(\varepsilon_2-\varepsilon_0)}
\frac{\gamma_{{\rm re},{\rm em} }^{-}\gamma_{{\rm re},{\rm ph} }^{+}\|\chi\|^2\|\psi\|^2+
\gamma_{{\rm re},{\rm em} }^{-}\gamma_{{\rm re},{\rm sink} }^{-}\|\chi\|^2+
e^{-(\beta_{\rm ph}-\beta_{\rm em})(\varepsilon_2-\varepsilon_0)}
\gamma_{{\rm re},{\rm ph} }^{-}\gamma_{{\rm re},{\rm sink} }^{-}\|\psi\|^2}
{\gamma_{{\rm re},{\rm em} }^{-}\gamma_{{\rm re},{\rm ph} }^{+}\|\chi\|^2\|\psi\|^2+
\gamma_{{\rm re},{\rm em} }^{-}\gamma_{{\rm re},{\rm sink} }^{-}\|\chi\|^2+
\gamma_{{\rm re},{\rm ph} }^{-}\gamma_{{\rm re},{\rm sink} }^{-}\|\psi\|^2}.
$$

Therefore for nonequilibrium environment the stationary state of the system will also be nonequilibrium.

Transfer rate of excitons to sink is equal (from the expression of the sink generator)
$$
F=2\gamma^{-}_{{\rm re,sink}}\rho_{11}-2\gamma^{+}_{{\rm re,sink}}\rho_{00}.
$$

For the considered here stationary state (\ref{rho22}), (\ref{rho11}), (\ref{rho00}), (\ref{Delta}) we get for the flow of excitons (taking in account $\beta_{\rm ph}=\beta_{\rm sink}$)
\begin{equation}\label{Flux}
F=\frac{2\|\chi\|^2\|\psi\|^2\left(\gamma_{{\rm re},{\rm em} }^{+}\gamma_{{\rm re},{\rm ph} }^{-}\gamma^{-}_{{\rm re,sink}}
-\gamma_{{\rm re},{\rm em} }^{-}\gamma_{{\rm re},{\rm ph} }^{+}\gamma^{+}_{{\rm re,sink}}\right)
}{\Delta}=
$$
$$
=\frac{2\|\chi\|^2\|\psi\|^2\gamma_{{\rm re},{\rm em} }^{-}\gamma_{{\rm re},{\rm ph} }^{+}\gamma^{+}_{{\rm re,sink}}
}{\Delta}\left(e^{\left(\beta_{\rm ph}-\beta_{\rm em}\right)(\varepsilon_2-\varepsilon_0)}-1\right).
\end{equation}

Let us discuss properties of the obtained expression for the flow of excitons.

1) Coefficient $\|\chi\|^2$ in the numerator describes the effect of superabsorption -- coherent amplification of absorption (the inverse effect to superradiance \cite{Dicke}). Possibility of superabsorption in photosynthetic systems was discussed in \cite{Superradiance}.

2) Analogously coefficient $\|\psi\|^2$ describes the effect of supertransfer -- coherent amplification of transfer. For discussion of possible effects of supertransfer in photosynthesis see \cite{tmf2014}, \cite{Superradiance}, \cite{Olaya-Johnson}, \cite{Mohseni1}, \cite{Mohseni2}.

3) The numerator of expression for the flow contains the product of three coefficients $\gamma$ (related to each of three reservoirs), and the denominator contains a linear combination of products of $\gamma$ for pairs of reservoirs. Thus the dependence of the flow on $\gamma_{{\rm re}, R}$ for $R={\rm em},{\rm ph},{\rm sink}$ has saturating form  --- for small $\gamma_{{\rm re},R}$ (corresponding to low intensity of the corresponding interaction, in particular for $R={\rm em}$ small $\gamma_{\rm re, em}$ corresponds to low intensity of light) the dependence of the flow on $\gamma_{{\rm re},R}$ will be linear and for high $\gamma_{{\rm re},R}$ this dependence will tend to constant.

4) When the state of environment tends to equilibrium, i.e. $\beta_{\rm em}\to \beta_{\rm ph}$, the flow tends to zero. This corresponds to absence of thermodynamic flows in equilibrium systems.

\section{Transfer rate for $\alpha\ne 0$}

In present Section we will generalize results of previous Section for the case $\alpha\ne 0$ (where $\alpha$ is the angle between bright photonic and phononic vectors ${\chi}$, ${\psi}$), i.e. in present Section these vectors $\chi$ and $\psi$ are non--parallel. For simplicity here we take Lamb shifts equal to zero $\gamma^{\pm}_{{\rm im}, R}=0$.

Let us consider expansion of (normed) bright photonic vector in components parallel and orthogonal to bright phononic vector 
$$
\widetilde{\chi}=\widetilde{\chi}_0+\widetilde{\chi}_1,\qquad \widetilde{\chi}_0 \| \widetilde{\psi},\quad \widetilde{\chi}_1 \bot \widetilde{\psi}
$$
$$
|\widetilde{\chi}_0\rangle=\langle\widetilde{\psi},\widetilde{\chi}\rangle |\widetilde{\psi}\rangle =|\widetilde{\psi}\rangle\langle\widetilde{\psi}| |\widetilde{\chi}\rangle,\qquad |\widetilde{\chi}_1\rangle=(1-|\widetilde{\psi}\rangle\langle\widetilde{\psi}|) |\widetilde{\chi}\rangle.
$$

Let us denote normed $\widetilde{\chi}_1$ by $\widetilde{\eta}$:
$$
\widetilde{\eta}=\frac{\widetilde{\chi}_1}{\|\widetilde{\chi}_1\|}.
$$
Hence $\widetilde{\eta}$ is orthogonal to $\widetilde{\psi}$ and lies in the same plane with $\widetilde{\chi}$ and $\widetilde{\psi}$, moreover \footnote{Actually the expressions for scalar products may also contain phase multipliers (modulus one complex numbers) but here we take these multipliers equal to one.}
$$
\langle\widetilde{\psi},\widetilde{\chi}\rangle=\cos\alpha, \qquad \langle\widetilde{\eta},\widetilde{\chi}\rangle=\sin\alpha,\qquad
|\widetilde{\chi}\rangle=\cos\alpha |\widetilde{\psi}\rangle + \sin\alpha |\widetilde{\eta}\rangle.
$$

Let us consider dynamics of the density matrix in the subspace of matrices of the form 
$$
\rho=\rho_{00}|0\rangle\langle 0|+\rho_{11}|1\rangle\langle 1|+\rho_{\psi\psi}|\widetilde{\psi}\rangle\langle \widetilde{\psi}|+
\rho_{\eta\eta}|\widetilde{\eta}\rangle\langle \widetilde{\eta}|+
\rho_{\psi \eta}|\widetilde{\psi}\rangle\langle \widetilde{\eta}|+\rho_{\eta\psi}|\widetilde{\eta}\rangle\langle \widetilde{\psi}|,
$$
\begin{equation}\label{norm}
\rho_{00}+\rho_{11}+\rho_{\psi\psi}+\rho_{\eta\eta}=1.
\end{equation}

This subspace of matrices is invariant with respect to dynamics generated by equation (\ref{total_gen}). Let us find in this space the stationary state for dynamics (\ref{total_gen}).

Equation for the stationary state for matrix element $\frac{d}{dt}\rho_{\psi\eta}=0$ implies expression for $\rho_{\psi\eta}$ in the stationary state (we consider the case where Lamb shifts are zero $\gamma^{\pm}_{{\rm im}, R}=0$)
$$
\rho_{\psi\eta}=\|\chi\|^2 \cos\alpha\sin\alpha\frac{ 2\gamma^{+}_{{\rm re,em}}\rho_{00}
-\gamma^{-}_{{\rm re,em}}\left(\rho_{\psi\psi}+\rho_{\eta\eta}\right)}
{\|\chi\|^2\gamma^{-}_{{\rm re,em}}
+\|\psi\|^2\gamma^{-}_{{\rm re,ph}}}.
$$

This value is real thus in the stationary state $\rho_{\eta\psi}=\rho_{\psi\eta}$.

Analogously in the stationary state (for $\alpha\ne 0$)
$$
\rho_{\eta\eta}=\frac{1}{\gamma^{-}_{{\rm re,em}}}\biggl[\gamma^{+}_{{\rm re,em}}\rho_{00}
-{\rm ctg}\,\alpha \gamma^{-}_{{\rm re,em}}\rho_{\psi\eta}\biggr].
$$

This implies the conditions 
$$
\rho_{\psi\eta}=\rho_{\eta\psi}=\frac{\|\chi\|^2 \cos\alpha\sin\alpha \biggl[ \gamma^{+}_{{\rm re,em}}\rho_{00}-\gamma^{-}_{{\rm re,em}}\rho_{\psi\psi}\biggr]}{\|\chi\|^2 \sin^2\alpha\gamma^{-}_{{\rm re,em}}  +\|\psi\|^2\gamma^{-}_{{\rm re,ph}}};
$$
\begin{equation}\label{etaeta}
\rho_{\eta\eta}=\frac{1}{\gamma^{-}_{{\rm re,em}}}
\frac{\gamma^{+}_{{\rm re,em}}\rho_{00}\left(\|\chi\|^2 (\sin^2\alpha-\cos^2\alpha)\gamma^{-}_{{\rm re,em}}  +\|\psi\|^2\gamma^{-}_{{\rm re,ph}}\right)+\|\chi\|^2 \cos^2\alpha (\gamma^{-}_{{\rm re,em}})^2
\rho_{\psi\psi}}
{\|\chi\|^2 \sin^2\alpha\gamma^{-}_{{\rm re,em}}  +\|\psi\|^2\gamma^{-}_{{\rm re,ph}}}.
\end{equation}

Let us investigate the conditions of stationarity for other elements of the density matrix taking into account above expressions for  $\rho_{\psi\eta}$, $\rho_{\eta\eta}$
$$
\frac{d}{dr}\rho_{\psi\psi}=0,\qquad \frac{d}{dr}\rho_{11}=0,\qquad \frac{d}{dr}\rho_{00}=0.
$$

We obtain
\begin{equation}\label{eq1}
-\rho_{\psi\psi}\bigg[
\|\chi\|^2 \gamma^{-}_{{\rm re,em}}\gamma^{-}_{{\rm re,ph}}+
\|\psi\|^2 (\gamma^{-}_{{\rm re,ph}})^2
\biggr]
$$
$$
+\rho_{11} \gamma^{+}_{{\rm re,ph}}\left({\|\chi\|^2 \sin^2\alpha\gamma^{-}_{{\rm re,em}}  +\|\psi\|^2\gamma^{-}_{{\rm re,ph}}}\right)
$$
$$
+\rho_{00}\|\chi\|^2 \cos^2\alpha \gamma^{+}_{{\rm re,em}}\gamma^{-}_{{\rm re,ph}}=0;
\end{equation}
\begin{equation}\label{eq2}
\rho_{\psi\psi}\|\psi\|^2 \gamma^{-}_{{\rm re,ph}}-\rho_{11}\left(\|\psi\|^2 \gamma^{+}_{{\rm re,ph}}+\gamma^{-}_{{\rm re,sink}}\right)+\rho_{00}\gamma^{+}_{{\rm re,sink}}=0;
\end{equation}
\begin{equation}\label{eq3}
\rho_{\psi\psi}\|\chi\|^2\|\psi\|^2 \cos^2\alpha \gamma^{-}_{{\rm re,em}}  \gamma^{-}_{{\rm re,ph}}+
$$
$$
+\rho_{11}\gamma^{-}_{{\rm re,sink}}\left(\|\chi\|^2 \sin^2\alpha\gamma^{-}_{{\rm re,em}}  +\|\psi\|^2\gamma^{-}_{{\rm re,ph}}\right)
$$
$$
-\rho_{00}\bigg[
\|\chi\|^2\|\psi\|^2 \cos^2\alpha \gamma^{+}_{{\rm re,em}}\gamma^{-}_{{\rm re,ph}}+
\gamma^{+}_{{\rm re,sink}}
\left(\|\chi\|^2 \sin^2\alpha\gamma^{-}_{{\rm re,em}} +\|\psi\|^2\gamma^{-}_{{\rm re,ph}}\right)
\biggr]=0.
\end{equation}

This and (\ref{etaeta}) imply
\begin{equation}\label{rho11-00}
\rho_{11}=\rho_{00}\frac{\|\chi\|^2\|\psi\|^2 \cos^2\alpha \gamma^{+}_{{\rm re,em}}\gamma^{-}_{{\rm re,ph}}+
\gamma^{+}_{{\rm re,sink}}\left(\|\chi\|^2 \gamma^{-}_{{\rm re,em}} +\|\psi\|^2\gamma^{-}_{{\rm re,ph}}\right)}{\|\chi\|^2\|\psi\|^2 \cos^2\alpha\gamma^{-}_{{\rm re,em}}\gamma^{+}_{{\rm re,ph}}+
\gamma^{-}_{{\rm re,sink}}\left(\|\chi\|^2 \gamma^{-}_{{\rm re,em}}  +\|\psi\|^2\gamma^{-}_{{\rm re,ph}}\right)};
\end{equation}
\begin{equation}\label{rho_psi_psi-00}
\rho_{\psi\psi}=\rho_{00}\frac{\gamma^{+}_{{\rm re,ph}}}{\gamma^{-}_{{\rm re,ph}}}
\frac{\|\chi\|^2\|\psi\|^2 \cos^2\alpha \gamma^{+}_{{\rm re,em}}\gamma^{-}_{{\rm re,ph}}+
\gamma^{+}_{{\rm re,sink}}\left(\|\chi\|^2 \sin^2\alpha\gamma^{-}_{{\rm re,em}} +\|\psi\|^2\gamma^{-}_{{\rm re,ph}}\right)}
{\|\chi\|^2\|\psi\|^2 \cos^2\alpha\gamma^{-}_{{\rm re,em}}\gamma^{+}_{{\rm re,ph}}+
\gamma^{-}_{{\rm re,sink}}\left(\|\chi\|^2 \gamma^{-}_{{\rm re,em}}  +\|\psi\|^2\gamma^{-}_{{\rm re,ph}}\right)};
\end{equation}
\begin{equation}\label{rho_eta_eta-00}
\rho_{\eta\eta}=\rho_{00}\frac{1}{\|\chi\|^2 \sin^2\alpha\gamma^{-}_{{\rm re,em}}  +\|\psi\|^2\gamma^{-}_{{\rm re,ph}}}
\biggl[\frac{\gamma^{+}_{{\rm re,em}}}{\gamma^{-}_{{\rm re,em}}}\left(\|\chi\|^2 (\sin^2\alpha-\cos^2\alpha)\gamma^{-}_{{\rm re,em}}  +\|\psi\|^2\gamma^{-}_{{\rm re,ph}}\right)+$$ $$+
\|\chi\|^2 \cos^2\alpha \gamma^{-}_{{\rm re,em}}
\frac{\gamma^{+}_{{\rm re,ph}}}{\gamma^{-}_{{\rm re,ph}}}
\frac{\|\chi\|^2\|\psi\|^2 \cos^2\alpha \gamma^{+}_{{\rm re,em}}\gamma^{-}_{{\rm re,ph}}+
\gamma^{+}_{{\rm re,sink}}\left(\|\chi\|^2 \sin^2\alpha\gamma^{-}_{{\rm re,em}} +\|\psi\|^2\gamma^{-}_{{\rm re,ph}}\right)}
{\|\chi\|^2\|\psi\|^2 \cos^2\alpha\gamma^{-}_{{\rm re,em}}\gamma^{+}_{{\rm re,ph}}+
\gamma^{-}_{{\rm re,sink}}\left(\|\chi\|^2 \gamma^{-}_{{\rm re,em}}  +\|\psi\|^2\gamma^{-}_{{\rm re,ph}}\right)}\biggr].
\end{equation}

Let us compute the transfer rate of excitons to the sink 
\begin{equation}\label{Flux1}
F=2\gamma^{-}_{{\rm re,sink}}\rho_{11}-2\gamma^{+}_{{\rm re,sink}}\rho_{00}=
$$
$$
=\rho_{00}\frac{2\|\chi\|^2\|\psi\|^2 \cos^2\alpha \left( \gamma^{+}_{{\rm re,em}}\gamma^{-}_{{\rm re,ph}}\gamma^{-}_{{\rm re,sink}}-\gamma^{-}_{{\rm re,em}}\gamma^{+}_{{\rm re,ph}}\gamma^{+}_{{\rm re,sink}}\right)}{\|\chi\|^2\|\psi\|^2 \cos^2\alpha\gamma^{-}_{{\rm re,em}}\gamma^{+}_{{\rm re,ph}}+
\gamma^{-}_{{\rm re,sink}}\left(\|\chi\|^2 \gamma^{-}_{{\rm re,em}}  +\|\psi\|^2\gamma^{-}_{{\rm re,ph}}\right)}=
$$
$$
=\rho_{00}\frac{2\|\chi\|^2\|\psi\|^2 \cos^2\alpha \gamma^{+}_{{\rm re,em}}\gamma^{-}_{{\rm re,ph}}\gamma^{-}_{{\rm re,sink}}}{\|\chi\|^2\|\psi\|^2 \cos^2\alpha\gamma^{-}_{{\rm re,em}}\gamma^{+}_{{\rm re,ph}}+
\gamma^{-}_{{\rm re,sink}}\left(\|\chi\|^2 \gamma^{-}_{{\rm re,em}}  +\|\psi\|^2\gamma^{-}_{{\rm re,ph}}\right)}\left(e^{\left(\beta_{\rm ph}-\beta_{\rm em}\right)(\varepsilon_2-\varepsilon_0)}-1\right).
\end{equation}

Matrix element $\rho_{00}$ (population of the lower energy level in the stationary state) can be computed using (\ref{rho11-00}), (\ref{rho_psi_psi-00}), (\ref{rho_eta_eta-00}) and density matrix normalization (\ref{norm}). Let us compare the obtained expression and formula  (\ref{Flux}) for the exciton flow for $\alpha=0$. Expression (\ref{Flux1}) for the flow contains $\cos^2\alpha$ in the numerator, in the first approximation this gives the dependence of the flow on angle $\alpha$ between bright photonic and phononic vectors $\chi$ and $\psi$.

\medskip

\noindent{\bf Remark.}\quad
In paper \cite{darkstates} possibility of manipulation of quantum states in analogous quantum system was considered. In this paper one of the reservoirs of nonequilibrium environment (one which describes interaction with light) contains also coherent component (corresponding to laser field). It was shown that in this case non--decaying ''dark'' states are generated, and these states can be observed by repeated interaction with coherent field (as in experiments with photonic echo, in particular photonic echo was observed in quantum photosynthesis). In this case the response for this interaction with coherent field was proportional to $\sin^2\alpha$ where $\alpha$ is the above mentioned angle between bright vectors.

In the present paper for the system under consideration we study the transfer rate in the stationary state in the case of non--zero angle between the bright states and investigate dependence of the transfer rate on this angle (in the first approximation the flow is proportional to $\cos^2\alpha$) and other parameters of the system. This dependence is the main result of the present paper.

Thus in the present model quantum transport and generation of non--decaying quantum ''dark'' states are competing processes (schematically this is described by the formula $\cos^2\alpha+\sin^2\alpha=1$).

Quantum transfer in the system interacting with nonequilibrium environment is the example of quantum thermodynamic machine, and transfer rate is the measure of effectiveness of this machine. Non--decaying quantum ''dark'' states are generated by leakage of of quantum states in the process of quantum transport.

\bigskip

\noindent{\bf Acknowledgments.}\quad
This work is supported by the Russian Science Foundation under grant 17-71-20154.

\end{document}